\RequirePackage{fix-cm} 
\documentclass{svjour3}                     
\smartqed  
\usepackage{graphicx}
\usepackage{latexsym,amsmath,natbib}
\usepackage[dvips]{color}
\usepackage{graphicx,txfonts}
\newcommand{\ms}{M$_{\odot}$}
\newcommand{\ls}{L$_{\odot}$}

\newcommand{\mpy}{M$_{\odot}$ yr$^{-1}$}
\newcommand{\macc}{$\dot{\rm M}_{acc}$} 
\newcommand{\mpe}{$\dot{\rm M}_{pe}$}

\newcommand{\lapprox}{$\stackrel {<}{_{\sim}}$}
\newcommand{\mdot}{{\it \.{M}}}
 \journalname{Space Science Reviews}
\newcommand{\aap}{{Astron. Astrophys.}}
\newcommand{\apj}{{Astrophys. J.}}

\begin{document}

\title{Disk Dispersal: Theoretical Understanding and Observational Constraints}
\titlerunning{Disk Dispersal}        
\author{U.~Gorti \and R.~Liseau \and Zs.~S\'andor \and C.~Clarke}
\institute{U. Gorti \at
              NASA Ames Research Center/SETI Institute, Mountain View, CA, USA \\
              Tel.: 001-650-810-0245, 001-650-604-3385\\
              \email{uma.gorti-1@nasa.gov}       
           \and
           R. Liseau \at
           Department of Earth and Space Sciences, Chalmers University of Technology, Onsala Space Observatory, SE-439 92 Onsala, Sweden
                         \and
           Zs.~S\'andor \at
             Konkoly Observatory of the Hungarian Academy of Sciences, H-1525 Budapest, P.O. Box 67., Hungary 
           \and
           C. Clarke \at
              Institute of Astronomy, University of Cambridge, Madingley Road,
               Cambridge, CB3 0HA
}
\date{Received: date / Accepted: date}
\maketitle
\begin{abstract}
Protoplanetary disks dissipate rapidly after the central star forms, on time-scales comparable to those inferred for planet formation. In order to allow the formation of planets, disks must survive the dispersive effects of UV and X-ray photoevaporation for at least a few Myr. Viscous accretion depletes significant amounts of the mass in gas and solids, while photoevaporative flows driven by internal and external irradiation remove most of the gas. A reasonably large fraction of the mass in solids and some gas get incorporated into planets. Here, we review our current understanding of disk evolution and dispersal,  and discuss how these might affect planet formation. We also discuss existing observational constraints on dispersal mechanisms and future directions.

\keywords{Protoplanetary disks\and Planet formation\and Accretion \and Winds} 
\end{abstract}
\section{Introduction}
\label{intro}
Disks provide the raw material for planet formation and the timescales on which they are dispersed therefore greatly affect their potential to form planets. Observations of disks, viz., the  cluster age-disk frequency plot, suggest that dust disk lifetimes are $\lesssim 3-5$ Myr (e.g., Haisch et al. 2001, Hernandez et al. 2007,  Ribas et al. 2015), and are only a fraction of the typical stellar lifetime. In such studies, which were initially conducted in the near-infrared (NIR), the fraction of young star cluster members with emission indicative of the presence of disk material is seen to dramatically decline with  cluster age. Although the reliability of the derived lifetimes is limited by the uncertainty in determining ages (e.g., Soderblom et al. 2014, Bell et al. 2013), there is clear evidence that NIR excess fractions decline with age from the classical T Tauri stage to the non-accreting, weak line T Tauri stage---a transition that spans a few Myr at the most.   Most disks seem to survive just long enough to allow planet formation (Lissauer et al. 2009), and only in a small fraction ($\sim$ 10\%, Winn \& Fabyrycky 2014) may the formation of giant planets be possible. If gas does manage to persist late into planet formation epochs, it can further affect planetary dynamics. Even small amounts of gas can influence the dynamics of young planetary systems: causing migration, damping eccentricities and mitigating the effects of planetesimal collisions. 

\begin{figure}[f]
\includegraphics[scale=0.7, clip]{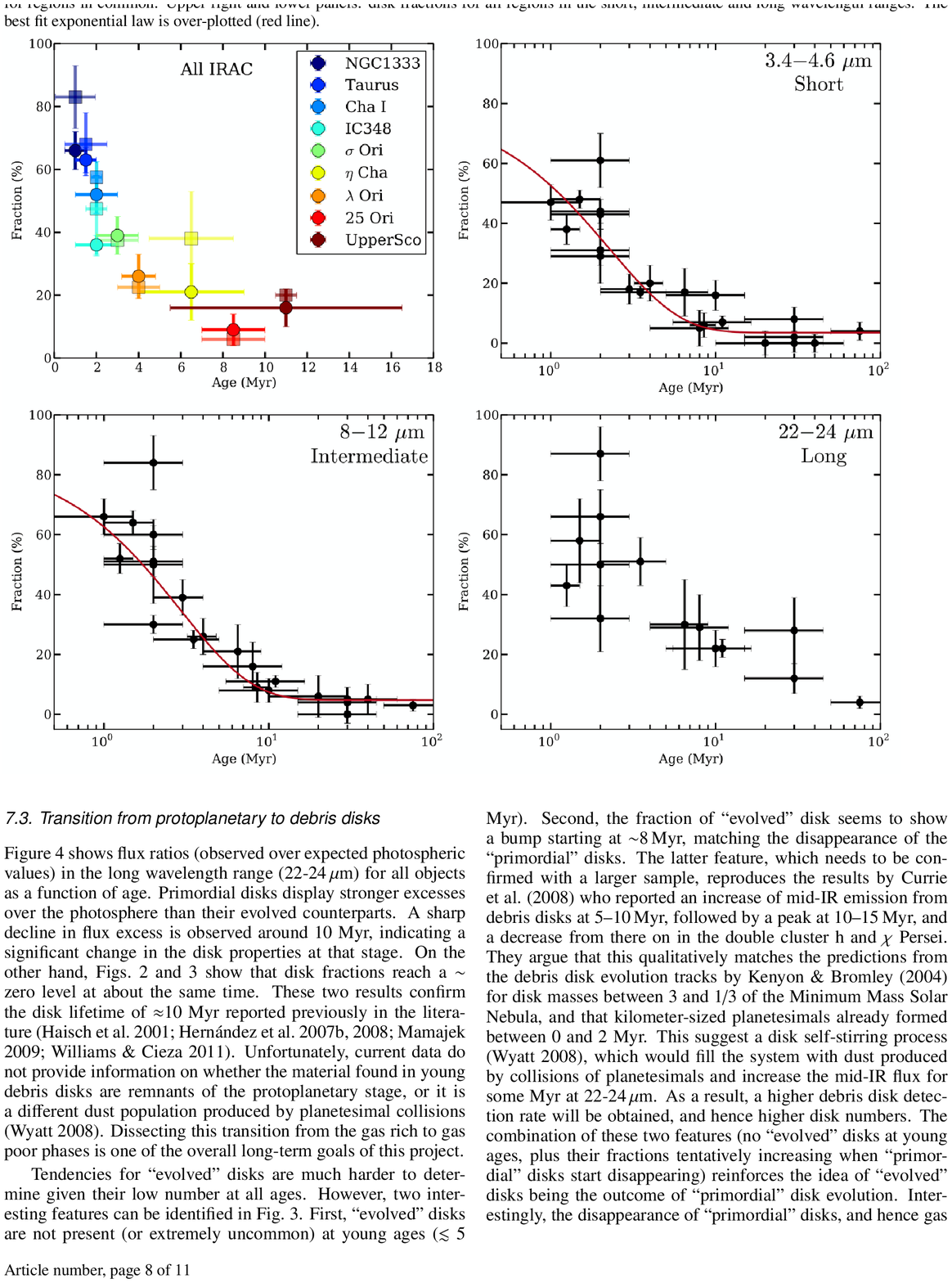}
\caption{ The disk frequency as a function of cluster age at different wavelengths as probed by {\em Spitzer} photometry. The inferred disk lifetime varies from $\sim 2-3$Myr at short wavelengths to $\sim 4-6$ Myr at longer wavelengths. (Figure from Ribas et al. 2015)}
\label{ribas14}
\end{figure}

Disk depletion lifetimes at longer wavelengths, tracing dust in regions farther from the star, are similar (e.g., see review by Williams \& Cieza 2011).  Sub-millimeter emission is rarely seen from disks without NIR excesses (Andrews \& Williams 2005), indicating either that the entire disk is depleted simultaneously or that the larger grains are lost earlier due to some combination of drift, fragmentation and/or planetesimal formation.  Dust at mid-infrared wavelengths appears to last slightly longer (e.g., Wahhaj et al. 2010, Hardy et al. 2015; see Figure~\ref{ribas14}), however, debris disks may contaminate emission statistics at these wavelengths. Nevertheless, the transition from optically thick to optically thin disks---believed to be represented by a class of objects called transition disks---appears to proceed from inside-out, .i.e., the inner dust is depleted first. Transition disks have lower accretion rates and dust holes in their inner radii  and constitute about $\sim 10$\% of the disk population (Williams \& Cieza 2011). While they are often believed to be a result of planet formation (e.g., Najita et al. 2015), these disks may represent a mixed class of objects  (see recent review by Espaillat et al. 2014) with only the lower mass disks on the verge of dispersal  (Ercolano et al. 2011, Owen \& Clarke 2012, Koepferl et al. 2013). 
 
Despite the fact that dust disk depletion timescales are fairly well known, disk dispersal mechanisms are not yet well constrained. In fact, the cluster age vs. disk frequency plots are also consistent with the interpretation that the $\sim 3-5$ Myr dust disk lifetime tracks the process of dust agglomeration into planetesimals. Planet formation from planetesimals, and even giant planet formation if the gas reservoir is still present, may then proceed over substantially longer periods.  In this interpretation, which holds if transition disks were mainly caused by planet formation, dust disk depletion does not necessarily imply dispersal of the disk material,  Gas, which dominates the disk mass through most of its evolution,  could in this case be removed on different timescales.  

However, observationally inferred timescales for the dispersal of gas in disks are less certain, mainly due to the fact that gas emission is intrinsically very faint. The earliest study was the seminal CO survey of  $\sim$ 10 disks by Zuckerman et al. (1995) which set a loose constraint on the gas disk dispersal time (at large $\sim 100$AU radii) of $\sim 10$ Myr. The FEPS legacy survey on the Spitzer Science Telescope, based on non-detections of H$_2$, set a dispersal timescale of the order of about $5-30$ Myr at radii $1-40$ AU (Pascucci et al. 2006). A more recent [OI]63$\mu$m survey using the Herschel Space Observatory (GASPS program) derived a similar timescale for the dispersal of gas ($\sim 5-200$AU). Dent et al.  (2013) quote that $\sim 18$\% of stars retain more than 1 M$_J$ worth of gas at ages of 4 Myr, and that all disks are dispersed by $\sim 10-20$ Myr.   Thus, gas dispersal times could, in principle, be longer than the $\sim3-5$ Myr dust disk depletion time. 

Different gas tracers have different sensitivity thresholds, making it difficult to compare gas and dust disk lifetimes.  In the inner disk ($\sim 1$AU), however, there appears to be clearer evidence of simultaneous dispersal. The fraction of accreting disks (with \macc$>10^{-11}$\mpy) in stellar groups declines on timescales similar to those of NIR excesses (Fedele et al. 2010), with some non-accreting sources (with \macc$<10^{-11}$\mpy) still showing IR excesses  (also Ingleby et al. 2013, Hardy et al. 2015). 
This may indicate that gas in the inner disk is removed first, consistent with dispersal scenarios (see Alexander et al. 2014). 

The main processes believed to disperse disks are a combination of viscous accretion and photoevaporation (see reviews by Hollenbach et al. 2000, Armitage 2011, Clarke 2011, Alexander et al. 2014) and to some extent, planet formation.   

Protostellar disks build the central star, hence it is to be expected that much of the disk gas is channeled into the central object. Evidence that disk evolution is largely driven by accretion during the bulk of the disk lifetime is provided by the fact that disk accretion rates approximately decline with age as expected from viscous accretion theory (e.g., Hartmann et al. 1998). To begin with, disks form due to the rotation of a gravitationally collapsing cloud core. Gravitational instabilities in the initially massive disk lead to strong accretion onto the central star (e.g., Laughlin \& Bodenheimer 1994).  At early stages,  magnetic fields  drive powerful jets and winds that carry away angular momentum from close to the star (e.g., K\"onigl 1991, Shu et al. 1994, see reviews by Li et al. 2014, Turner et al. 2014).  As star formation proceeds, infall decreases and the disk becomes gravitationally stable. Viscous accretion through the disk continues to transport mass inwards to the central star and angular momentum outwards. 

As the star reaches close to its eventual final mass, the rate of accretion declines (Hartmann et al. 1998, Mendigutia et al. 2012).  However, accretion does not proceed indefinitely and appears to abruptly halt with the disappearance of the inner gas and dust (e.g., Haisch et al. 2001, Fedele et al. 2010). Dispersal by pure accretion alone implies an indefinite expansion of the outer (gas) disk as angular momentum is re-distributed; there is no observational evidence to support such expansion. Further, relatively short ($\sim 10$ Myr) gas disk lifetimes call for an additional  dispersal  mechanism that removes gas from the system. 

Photoevaporation, whereby gas is heated to escape temperatures by the central star, is thought to disperse gas at later stages of evolution. Although observational evidence only requires the operation of some such mechanism at late stages, irradiation due to accretion-generated high energy photons should be even stronger and proceed to remove disk material even at earlier stages of disk evolution. When the magnetically-driven outflow/jet weakens along with the accretion rate (e.g., Hartmann et al.~1998,  White \& Hillenbrand 2004) and becomes more transparent, stellar high energy photons begin to penetrate the outflow column and irradiate the disk to heat its surface.  Penetration is first by hard X-ray\footnote[1]{Hard X-rays--$E\gtrsim$1keV; Soft X-rays--$\sim 0.1-0.3$keV; EUV--13.6eV-100eV; FUV-- 6-13.6eV. } and FUV photons when accretion rates fall to \macc$<10^{-6}$ M$_{\odot}$ yr$^{-1}$, and later by  EUV and soft X-ray  photons  (\macc$<10^{-8}$ M$_{\odot}$ yr$^{-1}$) (Gorti et al.~2009). This irradiation  marks the beginning of disk photoevaporation, where surface gas is heated to temperatures sufficient to overcome gravity and mass is lost in a slow, thermal wind.  Viscous accretion and photoevaporation subsequently work together to disperse the disk with time (Clarke et al. 2001). Theoretical estimates for a disk evolving under the influence of viscous accretion and photoevaporation,  including parametrized population synthesis models, agree with observationally inferred values ($\sim 1-10$ Myr) and can reproduce a range of observational diagnostics of
disk dispersal (e.g., Alexander et al. 2006, Gorti et al.~2009, Ercolano et al.~2009, Owen et al.~2010, 2012, Gorti et al. 2015).

The continuous removal of gas by photoevaporation may hold consequences for the evolution of the disk as it forms planets. While viscosity depletes gas mass uniformly at all radii and spreads the outer disk as it transports angular momentum, photevaporation preferentially removes gas at specific radii (inner $1-10$ AU and outer $\gtrsim 50$ AU, depending on which of EUV, FUV and X-rays dominate). If the mass loss rate due to photoevaporation (\mpe) is low, then planet formation proceeds unaffected by photoevaporation, except at late stages where the presence or absence of gas influences planet dynamics (e.g., Baruteau et al. 2014, Coleman \& Nelson 2015). If \mpe\ is high, then photoevaporation can influence early stages of planet formation by altering the gas/solids ratio (e.g., Gorti et al. 2015) and the type of planet formed (rocky vs. gaseous). 

Exoplanet statistics to date appear to indicate a relative paucity of gas giants (estimated frequency of $\sim 10$\% in solar-mass stars, Winn \& Fabrycky 2014), but an abundance of Super-Earths (M$_p \sim 3-10$M$_{\oplus}$) with some gas in an envelope. There must be insufficient gas present after the formation of planetesimals and rocky cores (a process thought to last $\sim 1$ Myr, Connelly et al. 2012) or gas giants would be more common. On the other hand, gas must necessarily be present at planetary core formation epochs to explain the frequency of Super-Earths. The rate at which gas is dispersed is thus closely aligned to planet formation timescales. 

 Photoevaporation may also influence the final architectures of planetary systems due to the migration of planets in a disk with gaps cleared during dispersal, leading to preferred semi-major axis distributions for exoplanets (Alexander \& Pascucci 2012, Ercolano \& Rosotti 2015).

This chapter, in keeping with the theme of this book, mainly deals with these above connections or links between disk dispersal and planet formation. We are interested not only in the lifetime of gas disks but also in the radial distribution of material at different stages of disk evolution. Dust evolution is discussed only as it pertains to gas dispersal mechanisms. (We refer the reader to chapters by Birnstiel et al. and Wyatt et al. for more in-depth reviews of dust.) We also do not discuss MHD winds which might deplete some disk mass especially at early stages. The structure of the chapter is as follows: we first describe early stages of disk evolution and accretion (\S 2), then photoevaporation due to the central star (\S3) and in a cluster environment (\S 4), planet formation and dispersal (\S 5), and in \S 6 describe observational constraints on disk dispersal.  We end with a discussion on future directions (\S 7). 

\section{Accretion}

It is widely accepted that the formation of stars and planetary systems is fundamentally governed by  the action of gravity and angular momentum. Whereas the overall picture is quite clear, the details are still far from fully understood. This concerns both theory and observation as both suffer from resolution problems. In addition, the highly complex interplay between physics and chemistry in the dusty plasmas of interstellar clouds leaves a complete description essentially intractable, at least for quite some  time to come.

We focus here on the formation  of low and intermediate- mass stars with masses \lapprox\,8\,\ms, for which Kelvin-Helmholtz time scales are longer than other time scales of relevance and consequently, for which we can follow the time evolution of the formation process. These less massive stars form in density enhancements of rotating molecular clouds. At supercritical density, the dense cores collapse, conserving angular momentum and forming flattened structures (Terebey et al. 1984, Walch et al. 2009, 2010) that eventually develop into disks and rings. 

\begin{figure}
\includegraphics[scale=0.4, angle=90]{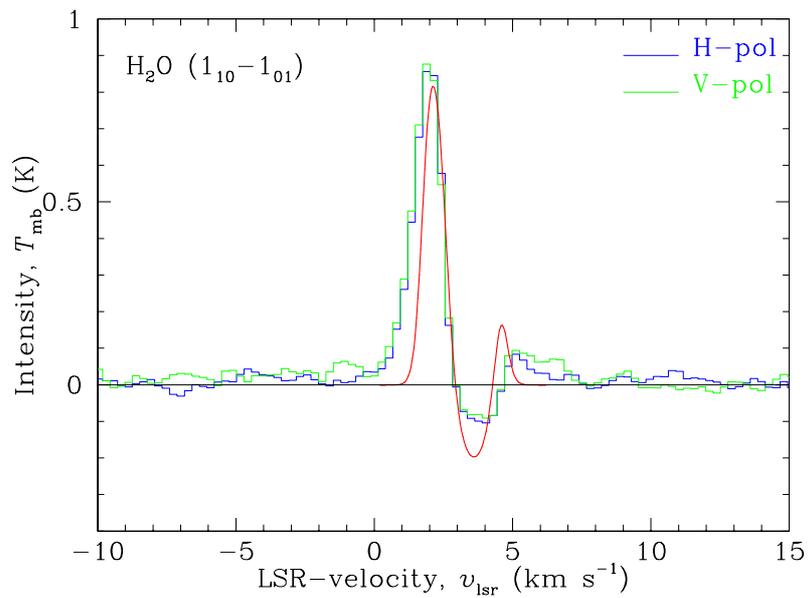} 
 \caption{Observed line profiles with {\it Herschel}-HIFI of the optically thick ground-state line of ortho-H$_2$O of a dense protostellar core, displaying, within the 38\"-beam (0.02\,pc), signs of both infall and outflow at the same time. All data are continuum subtracted around the zero-baseline, and the two polarizations are individually shown (blue and green) to demonstrate the high quality of the HIFI data. The red curve is a qualitative example from a radiative transfer model with center optical depth over a hundred, generating the central absorption. The blue-red asymmetry of the line core is due to the infalling gas in the unstable Bonnor-Ebert sphere, and the excess emission in the line wings is due to the outflow. }
 \label{infall}
\end{figure}

The hallmark of the dynamical phase of star formation, i.e. the infall phase, would be spectrally resolved molecular transitions with very high optical depths, as illustrated in Fig.\,\ref{infall}. There, the spectral infall signature is  qualitatively shown by a radiative transfer model, depicted as a smooth red line. Observed excess emission in the blue and red wings is attributed to outflowing gas. It is now firmly established that both early gravitational infall and later accretion are accompanied by mass loss phenomena, and these most often exhibit a bipolar geometry.

\subsection{Accretion processes in disks}
It is believed that the infall occurs onto the disk, and that the matter is accreted onto the central object through the disk. This, however, needs removal of a significant fraction of the angular momentum that is carried by the disk to prevent the break-up of the central object. Magnetic fields that thread through the core and the disk are invoked to act as a lever arm to brake the rotation. Outside magnetic dead zones, the fields are capable of providing the necessary viscosity due to the combination of magnetic and Reynolds stresses in a turbulent shearing flow (e.g, Balbus 2003, Cao \& Spruit 2013 and references therein).

The theoretical foundation of accretion disks was laid by the works of Shakura and Sunyaev (1973) and Lynden-Bell \& Pringle (1974). 
 The viscosity is described  parametrically by the product of a turbulent eddy size $H$ (of the order the pressure scale height) and its sound speed $c_{\rm s}$, i.e. $\nu=\alpha c_{\rm s} H$, where $\alpha$ is typically$10^{-4}$ to $10^{-2}$. The time evolution of the surface density ($\Sigma(r)$) of the disk is given by
 \begin{equation}
{{\partial \Sigma}\over{\partial t}} = {\frac{3}{r} }{{\partial}\over{\partial r}} \left(
\sqrt r {{\partial}\over{\partial r}} \left( \nu \Sigma \sqrt r \right) \right) 
\label{sdot}
 \end{equation}
 the solution of which describes the basic characteristics of an accretion disc, viz. that angular momentum is transported outward through the disk as matter is accreted into the inner regions. Disk observations have revealed rotational Keplerian signatures 
(Sargent et al. 1987, Dutrey et al. 1994, Olofsson et al. 2001, Guilloteau et al. 2014) 
  but predicted radial accretion drift velocities are too small to be measurable (on the order of some cm s$^{-1}$).  

Another observable would be the Spectral Energy Distribution (SED). The SED of a classical accretion disk is essentially that of a multi-temperature broadened blackbody. 
Recent models still exhibit these basic features (e.g., MCFOST, Pinte et al. 2006).
A useful quantity is  the integrated SED, i.e., the accretion luminosity $L_{\rm acc}= \eta\,G$ \mdot$_{\rm acc}\,(M/R)_{\rm star}$, where $0.5 \le \eta < 1$ is an energy conversion efficiency.  For typical parameters one finds $L_{\rm acc}=10$\,\ls, which is far above what had been determined from observations (Hartmann et al. 1997). 
 It was concluded that the accretion luminosity most likely is not steady in time, i.e. $dL_{\rm acc}/dt \propto d$\mdot$_{\rm acc}/dt \ne 0$, but variable within \mdot$_{\rm acc} \approx 10^{-8} - 10^{-4}$\,\mpy. The intermittent high states would be reached during FU\,Orionis type outbursts (Hartmann et al. 1996), whereas the low states would correspond to the typical T\,Tauri phase. Rise and decay times are of the order of 1\,yr and 100\,yr, respectively. Shorter time scales for $d$\mdot$_{\rm acc}/dt$ have been examined by
 Costigan et al. (2014). 

There are observational signatures of accretion. Optical emission lines from T\,Tauri stars, e.g. H$\alpha$, have been modeled as excited by shocks at the foot-prints of magnetized funnel flows (Muzerolle et al. 2001). However, it appears that the geometry and magnetic field topology are much more complex than envisaged in these one-dimensional models (e.g., see Fig.\,1 of Gunther 2013). 

\subsection{Mass loss accompanying accretion}

\paragraph{Jets}
Optically visible HH-objects (Haro 1950, Herbig et al. 1951, Reipurth et al. 2000)
and  jets  (e.g., Mundt et al. 1985, Shang et al. 2007)
are the cooling radiation from fast interstellar shock waves in star forming regions. Observations reveal jets on many length scales, viz. micro-jets (sub-arcsec) to pc-scales. As the name indicates, the collimation of jet flows is very high. The absence of detectable [O\,III] but prominent [S\,II] emission often indicates that the excitation (or density) is not very high, consistent with jet velocities not exceeding about 80 km s$^{-1}$. However, a number of jets are now known to emit in X-rays, implying jet velocities of the order of 500 km 
s$^{-1}$ or higher (
Liseau 2006 and references therein).
In many cases, but not all, these jets are seen together with generally much less collimated molecular outflows (Bally et al. 1983).  
\paragraph{Molecular outflows}

\begin{figure}
\includegraphics[scale=0.4]{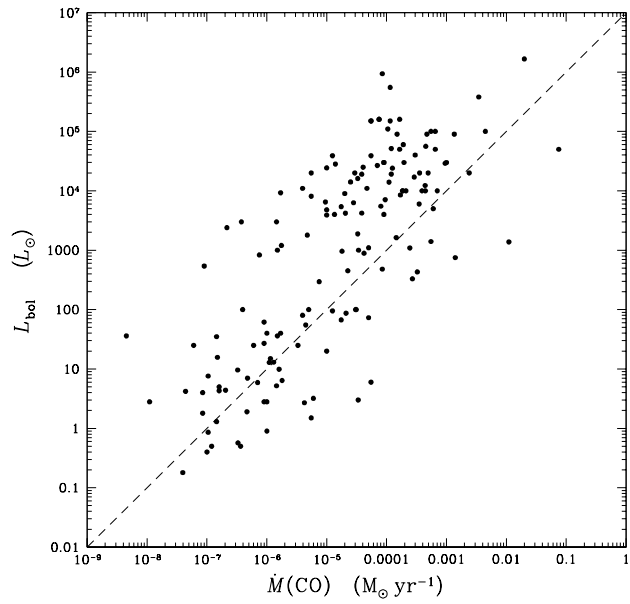}                        
  \caption{Mass loss rate, as determined from CO rotational line observations, viz. \mdot(CO), versus the bolometric luminosity $L_{\rm bol}$, of the outflow driving sources (data from Wu et al. 2004).   The dashed line has unit slope and is for reference only.}
 \label{CO-flows}
\end{figure}

Figure\,\ref{CO-flows} is based on the compilation of  literature data of CO-outflows by
Wu et al. (2004),  and shows the relation of the mass loss rate, as determined from CO-line mapping, and the bolometric luminosity of the driving sources, over seven orders of magnitude, and obtained from their infrared SEDs. The plot exhibits a large scatter that is due to the heterogeneity of the sample. However, in spite of this, it seems pretty clear that there is a dichotomy between low-luminosity ($< 100$\,\ls) and high-luminosity ($\ge 100$\,\ls) stellar sources. However, in both cases, the data can be fit by power laws, viz. $L_{\rm bol} \propto$ \mdot$_{\rm loss}^{\,a}$. While the low-luminosity distribution (where the luminosity is dominated by accretion) is consistent with $a = 1$, the distribution steepens at the higher end, with  $a = 2.5$  (see also Beuther et al. 2002 and references therein). 
 In the latter case, the luminosity most certainly is due to nuclear burning (objects already on the main-sequence). This power-law behavior strongly suggests that the underlying physics have common grounds and that the same physical laws govern these processes. 

Theories of jet acceleration all invoke the presence of relatively strong magnetic fields, whether for protostellar X-winds (Shu et al. 1994) or for disk-anchored disk winds (Pudritz et al. 2007, Li et al. 2014). The nomenclature ``wind'' describes the idea that the flows are initially poorly collimated. The precise nature of the interplay between disk-jet-molecular flow is difficult to determine observationally, primarily due to insufficient spatial resolution capability. 

However, there are a few clues: for instance, Hartigan et al. (1995) derived an outflow mass loss-to-accretion rate \mdot$_{\rm loss}$/\mdot$_{\rm acc}$\lapprox\,0.01  from optical observations, while White \& Hillenbrand (2004) derive a value $\sim 0.05-0.1$. Intruigingly, theoretical estimates of this parameter are 0.1\,\lapprox\,\mdot$_{\rm loss}$/\mdot$_{\rm acc} < 1$. Hartigan et al. (1995) concluded that these flows traced by [O I] forbidden lines may not carry enough momentum to drive the heavy CO-outflow. This was also the conclusion arrived at by Liseau et al. (2005)  in their detailed study of the protostellar object L1551-IRS, its jets and its CO flow.  For this young binary,  the dynamical mass is known (Rodriguez et al. 2003).   For the ratio of the rates, Liseau et al. (2005)
   found \mdot$_{\rm loss}$/\mdot$_{\rm acc}=0.23 \pm 0.10$ for the primary, and $0.7 \pm 0.3$ for the secondary, 
which were based on the large-scale CO outflow. These values are more in agreement with the theoretical prediction.
   
Since both the observed and theoretical values of the outflow mass loss rate are lower than the accretion rate, outflows cannot overwhelm accretion and hence do not play a major role in dispersing the disk at later stages.  

\section{Photoevaporation: Central Star}

\paragraph{Brief History} Photoevaporation was first studied in the context of massive stars by Hollenbach et al. (1994), who examined the effects of the rather strong radiation fields of massive stars on their disks (as suggested by Bally \& Scoville 1982). The basic premise is that the heating of the surface gas drives thermal winds from the disk ($c_s^2 > GM_*/r$) which then results in mass loss and a steady depletion of the disk material. Clarke et al. (2001) combined viscous evolution with photoevaporation to find that gaps open in disks at a preferred inner location (the gravitational radius, $r_g = GM_*/c_s^2$). Viscosity depletes matter interior to the gap, leading to inner holes. Adams et al. (2004) concluded that angular momentum support against gravity leads to the launching of flows at  smaller radii ($\sim 0.1-0.2 r_g$, also Begelman et al. 1983, Liffman 2003, Font et al. 2004). Alexander et al. (2006) recognized that the creation of a hole leads to the direct irradiation of the inner rim and results in a rapid dispersal of the outer disk. These theories were more recently extended to include the heating effects of  FUV and X-ray irradiation (Ercolano et al. 2009, Gorti \& Hollenbach 2009, Owen et al. 2012).  Overall, photoevaporation and viscous evolution together lead to the dispersal of gas on observed timescales ($\sim1-10$ Myrs).  For a more complete account of earlier work, we refer the reader to existing reviews of this topic (Hollenbach et al. 2000, Dullemond et al. 2007, Clarke 2011, Alexander et al. 2014).

\subsection{Overview of Photoevaporation}
The gravitational pull exerted by the central star decreases with distance, and so does the gas temperature; hence the ease with which flows can be launched depends on disk radius. The rate of change of surface density $\dot{\Sigma}_{pe}(r,t) = \rho_b v_{flow} \propto \rho_b c_s$, where $\rho_b$ is the density of gas at the base of the flow and the flow velocity $v_{flow}$ is proportional to the sound speed $c_s$ at this location.  $\dot{\Sigma}_{pe}$ is therefore sensitive to the density and temperature of the heated disk surface. In order to escape the system, the critical temperature needed at a given radius is $\sim 18,000/ r_{AU}$ K for a neutral atomic flow and $\sim 9500/r_{AU}$ K for fully ionized gas to escape with a flow velocity equal to the sound speed. Typical launch speeds are slightly subsonic, $\sim 0.5-1\ c_s$ (Owen et al. 2012, Gorti et al. 2015). 

\begin{figure}
\includegraphics[scale=0.5]{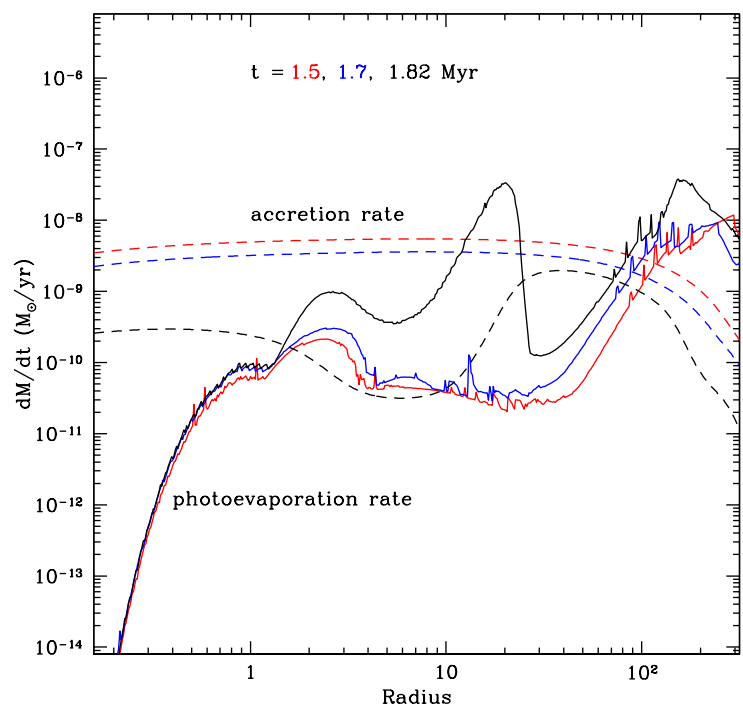}
\caption{ The increase in photoevaporation rate (solid lines) and the decrease in accretion rate (dashed lines) near the gap opening epoch is shown for a EUV+FUV+X-ray photoevaporation model. Here, the gap opens at $r\sim 10$AU, at about 1.8 Myr.   After gap opening, rim irradiation increases the photoevaporation rate, lowering $\Sigma$ and lowering the accretion rate further. Also note that photoevaporation rates are higher than the accretion rate in the outer disk($\gtrsim 100$AU) where significant mass loss occurs. }
\label{mdots}
\end{figure}

At early stages of evolution, accretion rates (\macc $\sim 3\pi \nu\Sigma$) are high compared to photoevaporative mass loss  (\mpe) and although photoevaporation may remove mass, its effects on the radial surface density distribution in the disk are minimal. As the surface density $\Sigma$ decreases with time due to viscous accretion, \macc\ declines  with it. \mpe, on the other hand, stays fairly constant with time; disk mass and its depletion is concentrated at the midplane whereas the density and temperature at the surface (and hence $\dot{\Sigma}_{pe}$) stay relatively unaffected. Eventually \macc\ drops below \mpe\ and this is when photoevaporation begins to play a dominant role in determining the evolution of the disk surface density with radius and time (Figure~\ref{mdots}). 

Gaps open in the inner disk when accretion can no longer replenish photoevaporative mass loss (e.g., Clarke et al. 2001). This happens at inner radii of $\sim 1-10$ AU for solar-mass stars due to the strong heating by the high energy photon flux. Gap opening halts the advection of mass from the outer disk, the inner disk drains rapidly and a hole is created. The disk continues to photoevaporate from the irradiated inner rim outward (e.g., Alexander et al. 2006). We note that holes can be created only if the high energy radiation field has a significant non-accretion generated component, i.e., it is mainly chromospheric/coronal in origin. If not, then the cessation of accretion chokes photoevaporation and gaps and holes cannot be sustained (see Gorti et al. 2015). In the case of FUV photoevaporation, flows are also launched from the outer disk where the surface temperature is still high but escape speeds are relatively low (e.g., Gorti  \& Hollenbach 2009, Owen et al. 2012). Under some conditions, gaps may also open in the outer disk. Since disks have shallow surface density distributions (e.g., Andrews et al. 2011), most of their mass is at large radii and photoevaporation here can affect the evolution of the entire disk. Viscous expansion in the outer disk is curtailed and the disk evolves into a shrinking torus of material (Gorti et al. 2009, 2015).

\begin{figure}
\label{diskevol}
\includegraphics[scale=0.5]{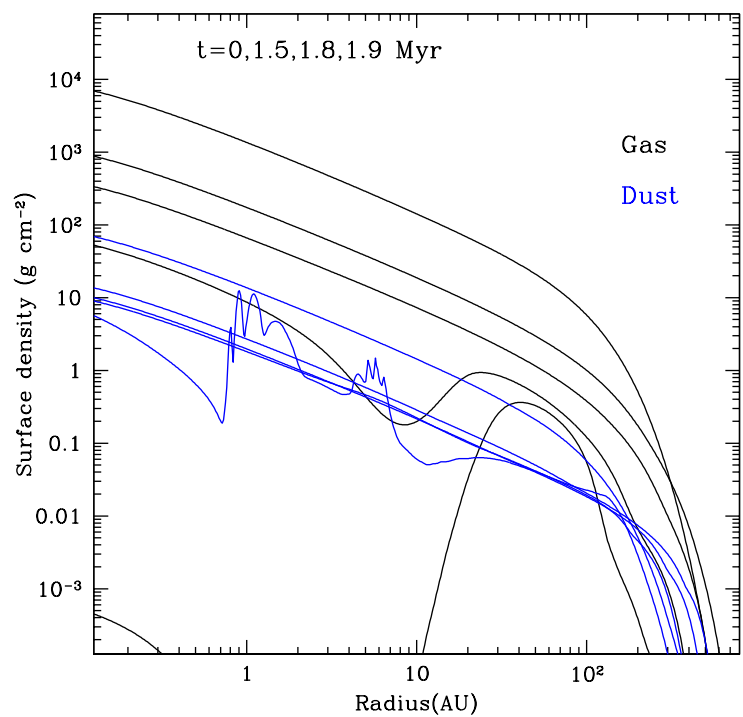}
\caption{The gas and dust surface density evolution of a viscously evolving disk, with FUV, EUV and X-ray photoevaporation, $M_d(0)=0.1$\ms, and $\alpha=0.01$. FUV photoevaporation leads to the creation of a gap at $\sim 3-10$AU and the gas disk disperses in $\sim 2$ Myr.  After the dispersal of the gas disk a substantial amount of dust is retained in the disk ($\sim 3\times10^{-4}$\ms). In these models, the largest solids are 1 cm in size, and no planetesimal formation is taken into account (see Gorti et al. 2015).  }
\end{figure}

Concurrent dust evolution plays an additional role for FUV photoevaporation. FUV heating of gas is due to collisions with energetic electrons ejected by small dust grains (small grains here include polycyclic aromatic hydrocarbons or PAHs) that absorb FUV photons (e.g., Bakes \& Tielens 1994).  The evolving abundance of small grains in the disk therefore affects the heating. However, small grains also attenuate FUV photons and their depletion increases penetration and shifts the base of the flow to higher densities. Since as noted earlier,  $\dot{\Sigma}_{pe}(r,t) \propto \rho_b \sqrt{T_{gas}}$, the depletion of small grains simultaneously decreases $T_{gas}$ and increases $\rho_b$ resulting in a smaller net effect on \mpe.  Overall, 2-fluid models of the evolution of the gas and dust (with a range of sizes) show that FUV mass loss rates are not significantly affected by dust evolution.  See Figure~\ref{diskevol} for the evolution of one such photoevaporating, viscous disk model.  Interestingly,  the gas/solids ratio in the disk is reduced by photoevaporation  (first noted by Throop \& Bally 2005) because dust grains are not coupled to the low-density gas in the wind and preferentially leave dust particles behind (Gorti et al. 2015).  Owen et al. (2011) describe how wind entrainment of dust could be observationally detected via their emission in edge-on disks.

\subsection{Photoevaporative mass loss rates}

As discussed in \S 1, the principal determinant of the relevance of photoevaporation for disk evolution and planet formation is the rate at which the disk loses mass. The mass loss rate also dictates how early on during evolution photoevaporation becomes important.  As long as accretion dominates (i.e., \macc\ $\gtrsim$ \mpe), viscous diffusion and advection will smear out any radial effects and replenish regions where photoevaporative mass loss has occurred. 

While there is reasonable agreement on the qualitative behaviour of disk photoevaporation between different models, the calculated mass loss rates vary by over two orders of magnitude, from $10^{-8}-10^{-10}$ \mpy. At the high end, \mpe\ $>$\macc\ during the Class II stage and photoevaporation determines the radial distribution of disk material and can significantly affect planet formation. Rapid dispersal may even preclude the formation of planets. For low \mpe, the role of photoevaporation may be limited to clearing the disk of small amounts of remnant gas and facilitating the circularization of planetary orbits (Kominami \& Ida 2002). 

Some of the differences in estimated \mpe\ can be attributed to the high energy photons under consideration. For pure EUV models, \mpe\ can be low  ($ < 10^{-10}$ \mpy). Although ionized gas is heated to $\sim 10^4$ K, EUV is absorbed at very small column densities and the low $\rho_b$ results in low mass loss rates.  High EUV luminosities can yield higher mass loss rates, but recent studies suggest photon luminosities $\lesssim 10^{40}-10^{41}$ s$^{-1}$ ($L_{EUV} \sim 10^{30}$ erg s$^{-1}$) and hence that the associated \mpe\ is low (Pascucci et al. 2014). 
 
X-ray and FUV models result in higher mass loss rates, $\sim 10^{-8}-10^{-9}$ \mpy\ for typical stellar radiation fields. The calculated \mpe\ is in general sensitive to the density and temperature structure of the disk which now has to be determined unlike in the EUV case. The disk structure is in turn based on disk chemistry and calculated cooling rates that are all highly model-dependent (e.g., see Rollig et al. 2007). 
For X-ray photoevaporation models, Owen et al. (2012)  however state that the resulting flow properties are insensitive to the detailed thermal and density structure of the upper disk layers but are instead set by a criticality condition at height $\sim R$ above the disk plane where the flow makes a subsonic to supersonic transition. They further argue that provided the flow structure is optically thin to the X-rays dominating the heating at this surface, the mass loss rate is independent of any complex thermochemical effects at greater depth in the flow. If this condition is not met however, then the flow instead makes a sonic transition in regions where heating is dominated by FUV and hard X-rays and then it is essential to calculate the disk vertical structure.

The X-ray spectrum assumed also impacts disk temperatures and hence \mpe (Gorti et al. 2009); Gorti, Hollenbach et al. assume that soft X-rays ($0.1-0.3$keV) are mostly absorbed in accretion and outflow columns before they reach the disk surface, while Ercolano, Owen et al. 
 assume a small covering factor of the accretion columns and no absorption in the column, to attain much higher temperatures in their disk models. The latter do not consider molecular cooling, but Gorti et al. find that molecular cooling can be important for regions penetrated by hard X-rays ($\gtrsim 1$keV) and their model disks have cooler temperatures.

Gorti et al. further treat the flow dynamics using simple analytical estimates drawn from previous work on thermal winds (Begelman et al. 1983, Liffman 2003, Adams et al. 2004, Waters \& Proga 2014), but conduct detailed thermo-chemical modeling. Owen et al. claim that the flow structure is unimportant for soft X-rays, and adopt the opposite approach to solve for \mpe\ using full radiation hydrodynamics models with simpler thermal physics.  However, Gorti et al. include FUV photoevaporation along with X-rays and EUV, and in spite of a smaller role for XEUV photons, get comparable mass loss rates. The high mass loss is partly due to the time-dependent accretion FUV luminosity, which can be substantial in disks (e.g., Gullbring et al. 1998). 
More recent models by Owen et al. (2012) also find that FUV photoevaporation can dominate if the FUV luminosities are high and better reconcile the differences between the two groups. 

Disk mass loss rates can vary depending on a number of parameters, e.g. stellar mass, initial disk mass and radius, viscosity in the disk, EUV, FUV and X-ray luminosities, and the time-dependent XEFUV spectrum (e.g., Ercolano et al. 2009, Gorti et al. 2009, Owen et al. 2012, Gorti et al. 2015), many of which are known to vary widely---often by an order of magnitude or more---in young stars.  This diversity results in photoevaporation rates that can vary widely depending on the system, and \mpe\ can generally range from $10^{-11}$ to $10^{-7}$ \mpy\ for $M_*\sim 0.1-3$\ms. 

The disk lifetime for a disk of initial mass $M_d(0)$ and a time-averaged photoevaporation rate
$\langle$\mpe$\rangle$ can be  approximately  written as
\begin{equation}
\tau_{disk} \approx 10^7 \left(\frac{M_d(0)}{0.1{\rm M}_{\odot}}\right)^{2/3} 
\left(\frac{\langle\dot{M}_{pe}\rangle}{10^{-9}{\rm M}_{\odot}\ {\rm yr}^{-1}}\right)^{-2/3} \quad {\rm yr},
\label{taudisk}
\end{equation}
for a linear viscosity profile and assuming $\alpha=0.01$ (e.g. Clarke et al. 2001,  Gorti et al. 2015). For the fiducial photoevaporation rates discussed above of $10^{-8}$ to $10^{-9}$ \mpy  and an initial disk mass of 0.1\ms, the corresponding disk lifetimes are thus $\sim 2-10$ Myr. We note that, in principle, \mpe\ can change with time as the disk evolves, and the equation above represents an average rate over the disk lifetime (see Gorti et al. 2015).

\section{Photoevaporation in the Cluster Environment }
So far we have considered three flavours of disc photoevaporation driven by the EUV, FUV or X-ray radiation from the disk's central star.  In the crowded environments of young clusters, however, there is also the possibility of ÔexternalÕ photoevaporation by the radiation field produced by (more massive) neighbouring stars. This is particularly to be expected in the case of the EUV and FUV where stars' photospheric outputs are a strong function of stellar mass (Diaz-Miller et al. 1998) and where, even taking into account the relative rarity of higher mass stars, the integrated contribution to the EUV and FUV backgrounds peaks at masses in the range 10--55\ms\ (Fatuzzo \& Adams 2008). This last point is important in assessing the types of cluster environments in which one expects external photoevaporation to be important. At high cluster membership number (N), even the top end of the IMF is statistically well populated and thus the distribution of total UV luminosity at a given N is sharply peaked at a value that simply scales with N. In the case of low N clusters, by contrast, there are large stochastic variations in the population of the upper IMF and the distribution of UV luminosities at given N is broad with a median that is well below the mean. This just means that external photoevaporation is unimportant in low N clusters, partly because the over-all number of stars is lower and partly because, in consequence, the IMF often ends up not containing the most massive stars that dominate the UV budget: see Fatuzzo \& Adams 2008 for a detailed analysis of this issue. The behaviour of the X-ray background is more complex because X-ray luminosity does not increase monotonically with stellar mass, attaining a minimum in the case of fully radiative stars in the A star range. Measurements of diffuse X-ray emission in rich clusters such as M17 or the Rosette Nebula (e.g. Townsley et al. 2003) suggest that this emission is largely dominated by early type O stars (Feigelson et al. 2003) since the ONC (for which the most massive star is of spectral type O5) lacks a comparable diffuse field.

\subsection{External X-ray photoevaporation}
It is easy to demonstrate that external X-ray photo-evaporation is negligible compared with internal X-ray photo-evaporation. The X-ray driven mass loss rate at each radius scales linearly with the X-ray flux (since, in the ionisation parameter formulation, the density corresponding to the local escape temperature is proportional to $F_X$). Even the X-ray flux reported in the ONC near $\theta_1$C Ori is orders of magnitude less than the X-ray flux of an average T Tauri star at a radius of 100 A.U. We therefore do not consider this possibility further.
\subsection{EUV + FUV photoevaporation from a single star}
We now consider the simplest case where the external UV field in a cluster is dominated by a single star. This is approximately the case in the ONC, where $\theta_1$C Ori has an ionising output that substantially exceeds that of the other O stars in the cluster core. We will not for now concern ourselves with the fact that there are other O stars such as 
$\theta_1$A or $\theta_1$B that are the major contributors of FUV flux to their nearest neighbours.
If we consider the luminosity from a single source, rather than an isotropic background, then we do not expect the resulting wind structures to be spherically symmetric and indeed the photoevaporating discs in the ONC show generally cometary morphologies, often being brighter on the side facing $\theta_1$C and with a tail pointing away from this star (Tsamis et al. 2013). Detailed modeling of proplyd morphology assumes that the hemisphere of the ionised flow is directly illuminated whereas the far-side receives a diffuse EUV field derived from recombinations in the nebula (see Henney \& Arthur 1998). We will however follow Johnstone et al. (1998) in setting out a simplified spherically symmetric photoevaporation model, an approach that yields mass loss rates that agree to order unity with more complex modeling.

EUV radiation impinges on low density gas above the disc and sets up an ionisation front in which the integrated number of recombinations per unit area matches the starÕs ionising flux (one can readily justify {\it a posteriori} the neglect of additional consumption of EUV photons in ionising neutral material flowing through the ionisation front since this turns out to be a small addition for typical flow parameters; Churchwell et al. 1987). The ionisation front represents a contact discontinuity of the flow for which imposition of mass and momentum conservation on either side results in a flow that is transonic in the ionised region (i.e. the ionised gas flows out at around 10 km s$^{-1}$) but enters the ionisation front sub-sonically. Naturally there must be heating processes ({\em not} involving ionising photons) that produce the pressure gradients throughout the neutral region, but provided this flow is subsonic throughout the neutral region it is in causal contact with the ionisation front and it is thus conditions at the ionisation front that set the mass flow rate. This situation is known as EUV photoevaporation, even though FUV heating must contribute also in setting up the neutral flow that feeds the ionisation front. There is however a qualitatively different situation if the solution contains a sonic point in the neutral
flow. In this case the flow has to undergo shock deceleration before it can enter the ionisation front and thus neutral flow below the shock is causally decoupled from the ionisation front. This means that the agent heating the neutral flow (the FUV) now {\em sets} the wind mass loss rate.

For EUV driven flows, ionisation equilibrium at the ionisation front implies that $\int n^2dr$ scales with the ionising flux so that, for a spherical transonic flow one obtains:
\begin{equation}
\dot{M}_{EUV} = 10^{-8} \Phi_{49}^{1/2} d_{17}^{-1} r_{d14}^{3/2} \quad {\rm\ M}_{\odot} {\rm yr}^{-1}
\label{meuv}
\end{equation}
(Johnstone et al. 1998) where $\Phi_{49}$ is the ionising luminosity of $\theta_1$C in units of $10^{49}$ s$^{-1}$, $d_{17}$ is the distance of the disc from $\theta_1$C in units of $10^{17}$ cm and $r_{d14}$ is the radius of the ionisation front in units of $10^{14}$ cm. Note that in the case of EUV driven winds (where by definition the neutral flow is thin and sub-sonic), $r_{d14}$ roughly equates with the disc radius.

In the case of FUV flows, by contrast, the wind mass flux is set by the maximum density of the flow for which FUV heating is effective. For the strong FUV fields in the centre of Orion, this condition is set by dust absorption (rather than molecular self-shielding) and therefore imposes a maximum column in the FUV heated neutral flow. Since the scale height of FUV heated gas in the outer disc is of order $r_d$ (i.e. the FUV heats to around the escape temperature of the outer disc), the number density in the flow is set by $n r_d \sim 10^{21}$ cm$^{-2}$. Putting this together (again assuming spherical symmetry and that the flow velocity is transonic in the neutral gas, i.e. with velocity $\sim 3$ km s$^{-1}$) we obtain the relation:
\begin{equation}
\dot{M}_{FUV} = 10^{-8}  r_{d14} {\rm\ M}_{\odot} {\rm yr}^{-1}
\end{equation}
(Johnstone et al. 1998). Note that, unlike, the expression for $\dot{M}_{EUV}$, this rate does not depend explicitly on the strength of the FUV field. However, this expression does not apply at arbitrarily low FUV levels because at some point self-shielding by molecular hydrogen becomes more important than dust absorption in setting the column of neutral gas (also see  discussion in \S 4.5). Storzer \& Hollenbach (1999) argued that the critical flux level was $G_0 \sim 5\times10^4$ which corresponds to distances from $\theta_1$C of around 0.3 pc.
\footnote{G$_0$ is the ratio of the FUV flux to its ambient value
in the solar neighbourhood which is $1.6 \times 10^{-3}$ erg cm$^{-2}$ s$^{-1}$
(Habing 1998).}

We thus have a schematic picture which would imply three radial zones with distinct disc photoevaporation properties. Noting that $\dot{M}_{EUV}$ scales inversly with distance from
$\theta_1$C while $\dot{M}_{FUV}$	is constant over a range extending out to a limiting $G_0$, we then have i) a small inner region with very fierce EUV photoevaporation, ii) an intermediate FUV zone with spatialy constant mass loss rate and iii) an outer EUV zone which resumes at the point that FUV photoevaporation is no longer effective. The latter is at $\sim$ 0.3pc while the interface between i) and ii) is set by equality of the EUV and FUV mass loss rates at that point, implying
a radius of $\sim0.15 r_{d14}^{1/2}$.

\subsection{The observational evidence for external photoevaporation in Orion}
An immediate implication of the FUV model described above is that the ionisation front is spatially offset from the disc because it is separated from the disc by a spatially thick supersonic neutral flow. Such a structure corresponds well to the `proplyds' first imaged in Orion by O'Dell et al. 1993 (see Fig.). (Note that whereas O'Dell applied the term `proplyd' to all the circumstellar structures imaged in Orion, including the pure silhouette discs, we will here restrict the definition to those showing offset ionisation fronts, regardless of whether a central silhouette disc is also detected.) The spatial distribution of the bulk of the proplyds accords well with the ÔFUV zoneÕ described above. (Note that the innermost EUV zone is plausibly filled in by projection effects).

Although the bulk of the proplyds indeed lie within the 0.3pc radius predicted above, there are several tens of proplyds at larger radius, only some of which are explicable as being instead powered by the more modest FUV heating provided by $\theta_1$A and B (Vicente \& Alves 2005). It is possible that FUV winds may be driven at lower values of $G_0$ than argued above but in this case it is unclear why the number of proplyds at large radii is small, though non-zero. Alternatively, Clarke \& Owen (2015) suggested that these far flung prolyds result from external EUV ionisation of a neutral flow driven by internal X-ray photoevaporation. They showed that the numbers and sizes of far flung proplyds are consistent with statistical expectations based on the X-ray luminosity function (though the correlation between instantaneous X-ray luminosity and resolvable proplyd structure on a source by source basis is weak, an effect that might be attributed to variability). Similarly there are far flung proplyds detected in Carina (Smith et al. 2003), NGC 3603 (Brandner et al. 2000) and Cyg OB2 (Wright et al. 2012) although some of these are likely to instead be ionised clumps of molecular gas (Sahai et al. 2012 a,b).

Having established that theory more or less correctly predicts the spatial distribution of proplyds, we now turn to the mass loss rates in these objects. These were first deduced from resolved radio free-free observations (Churchwell et al. 1987) combined with the assumption of transonic expansion in the ionised flow.	As	the	resulting	flow	rates	(comparable	with	 $\dot{M}_{FUV}$	as	given	above)	imply	 problematically	short disc lifetimes (see below) O'Dell (1998) instead suggested that the structures might be pressure confined. These high mass loss rates were however subsequently confirmed through emission line modeling (Henney \& OÕDell 1999) which showed that the kinematics were clearly incompatible with a static, pressure confined structure but involved free expansion.

These high mass loss rates, both predicted and observed, imply that the total mass photoevaporated over the cluster lifetime is of order a solar mass. This is far more than the initial gas reservoir in discs and implies that we would expect to see a distinct deficit of discs at the present epoch in the centre of Orion. The most troubling aspect of the Ôproplyd lifetime problemÕ is that the disc fraction in the core of Orion is actually high; in fact, when projection effects are taken into account, the disc fraction is $\sim$100\% within the central FUV region. St\"{o}rzer \& Hollenbach (1999) suggested that the unexpected survival of discs at small radii could be explained if stars were on very radial orbits and just `lit up' as proplyds during their brief sorties through the core region. However, such an orbital configuration was found to be unsustainable in N-body simulations (Scally \& Clarke 2001).  Another effect that may mitigate the very short predicted survival times is that FUV photoevaporation is considerably less efficient in the case of ``sub-critical" discs (i.e., those in which the escape velocity at the disc outer edge exceeds the sound speed in the heated gas). Adams et al. (2004) have shown that the mass loss rate in this case declines with decreasing radius much more steeply than implied by Eq.(4); although Clarke (2007) suggested that this might enhance the survivability of small disks in the ONC, the important of this effect is limited by the fact that even in the case of negligible FUV mass loss, the rate reverts to the EUV rate (Eq.(3)).

This then only leaves two possibilities (or a combination of the two): unexpectedly large disc masses and/or a recent Ôswitch-onÕ time for $\theta_1$1C.

Large disc masses are however not indicated by any of the sub-mm studies in Orion (see Mundy et al. 1997, Bally et al. 1998, Eisner \& Carpenter 2006, Mann et al. 2014) and although estimation of gas disc masses is traditionally beset by systematic uncertainties regarding the gas to dust ratio and grain opacity, it is clear that in a comparative sense the discs in Orion are not unusually massive. This instead encourages the Ôlast gaspÕ interpretation, in which $\theta_1$C has only been `switched on' (or perhaps, more correctly, been optically revealed) over a timescale that is a small fraction of the cluster lifetime. In this scenario, the Orion proplyds (and indeed all disc emission in the inner parts of Orion) will be gone within a timescale of $>10^5$ years (also see Clarke 2007).

Although it is slightly uncomfortable to argue that we are witnessing our nearest massive star formation region at a Ôspecial epochÕ there is some circumstantial evidence that this is indeed the case. The disc mass distribution derived by Mann et al. (2014) does show some hints that preferential depletion of disc mass is indeed starting to occur at the very small radii characterising the inner EUV zone (where the mass loss rates are even higher than in the FUV zone: Eq.~\ref{meuv}). Secondly, proplyds indeed seem to be rather rare in other star forming regions (Yusef-Zadeh et al. 2005, Balog et al. 2006, Koenig et al. 2008), although this result has to be interpreted with caution given that resolving prolyds is more challenging in more distant environments than in Orion. Perhaps the most conclusive argument indicating that Orion is being observed at a special epoch is that it is unusual among massive clusters in not showing a deficit of disc bearing stars in the proximity of massive stars. It is to the disc demographics in other clusters that we now turn.

\subsection{The distribution of disc bearing stars in massive star clusters}
A number of studies have now conducted disc censuses in star forming regions where one can reasonably expect that external photoevaporation will be important: for example Guarcello et al. (2007, 2009, 2010) in NGC 6611, Fang et al. (2012) in Pismis 24 and Guarcello et al. (2015) in Cygnus OB2. In all cases it is seen that the disc fraction is lower in the proximity of massive stars. Guarcello et al. (2015) have quantified this effect by generating an estimated map of the ambient FUV field across Cyg OB2 and have demonstrated that the disc fraction declines monotonically with increasing FUV flux, being around a factor two lower in the highest FUV versus lowest FUV category. There is also some evidence from this analysis that the (anti-)correlation between disc frequency and FUV flux is more convincing than that between disc frequency and the expected frequency of star-disc collisions. This is to be expected given that in any environment where star-disc collisions are at all significant, the effect of external photoevaporation is likely to be much more important (Scally et al. 2001).

\subsection{The effect of mild FUV fields in sparse star forming environments}
Environments like the ONC are highly atypical in terms of the strength of the ambient FUV field.
The population synthesis exercise of Fatuzzo \& Adams (2008), which assembled clusters according to the observationally inferred spectrum of cluster richness, demonstrated that regions which - like the core of Orion - have $G_0$ values in excess of $10^4$ are environments in which at most a few percent of star formation occurs. On the other hand, other star forming environments span an enormous dynamic range in $G_0$ values ($4-6$ orders of magnitude). Adams et al. (2004) studied flow solutions in a range of low $G_0$ environments, arguing that in this case photoevaporation is predominantly in the (cylindrical) radial direction from the outer edge of the disc. Recently Facchini et al. (Facchini, S., Clarke, C.,  Bisbas, T. 2015, submitted to MNRAS) have revisited the problem and obtained solutions over a wider range of parameter space, additionally iterating on the thermal solution to take account of the fact that only small grains are entrained in the flow. Preliminary solutions indicate rather significant mass fluxes for discs larger than 100 A.U. even at low $G_0$. A generic feature of these solutions is a cliff in the gas surface density at the disc outer edge and then a low density plateau of (nearly dust free) gas at larger radii. Such solutions offer the prospect of possibily constraining the properties of such flows through deep molecular line imaging.

Regardless of the numbers that emerge from this exercise, it is also worth noticing that the interplay between viscosity and outer edge can have some unexpected outcomes. A viscous disc without a photoevaporative flow evolves towards a viscous similarity solution in which the disc outer edge grows in such a way that the viscous time is always of order the disc age. Consequently the surface density in such a disc declines as a power law in time and - as the viscous time gets longer and longer as the disc expands - the time required for the disc to clear (e.g. become optically thin in the near infrared) is extremely long.

It might be thought that the addition of a disc flow from the outer edge might at best reduce the disc lifetime to a value equal to $t_{evap}$, the ratio of the present disc mass to the photoevaporation rate. In fact, however, the effect is much more dramatic: when external photoevaporation is coupled to the viscous flow then the disc at some stage stops growing and then shrinks from the outside in (an effect also seen in the internal FUV photoevaporation models of Gorti et al. (2009), Gorti et al. (2015)). In this case, the viscous timescale decreases with time and hence the clearing is accelerated. This effect results in the disc clearing an order of magnitude faster than $t_w$ (Clarke 2007, Anderson et al. 2013) and it is obvious that in this case most of the disc mass is cleared by accretion rather than photoevaporation. Nevertheless photoevaporation is playing a decisive role in preventing disc spreading and thus keeping the viscous timescale short.

\section{Planet formation and photoevaporation}
\subsection{Inventory of disk's solid material and its depletion}
For disks that form planetary systems, the planets themselves form an important sink for the initial mass in solids. If giant planets are present, then a fraction of the disk gas is also consumed. In our solar system, there was a minimum of about $\sim 1-3\times 10^{-4}$ \ms\ of solids and a little more than a Jupiter mass or $\sim 1-3\times 10^{-3}$\ms\ of gas that  was ``depleted" into planet formation. We use the term ``depletion" because this disk component is dark for most external systems. A disk-less object may very well have a nascent planetary system that is undetected. The protosolar disk therefore lost 10 times as much gas as the dust, and this is a fact that dispersal theories must necessarily explain. 

Exoplanetary disks, especially the birth-sites of the compact Kepler systems, may efficiently convert even more mass into rocky planets (e.g., the Minimum Mass Extrasolar Nebula models of Hansen \& Murray 2012, Chiang \& Laughlin 2013). The low frequency of gas giants ($\lesssim 5-10$\%, Winn \& Fabrycky 2014) and the abundance of Super-Earths ($\sim $50\%) indicates that in most disks gas is depleted relative to dust even more than in the solar nebula. 

Unless migration is not as efficient as theories predict, accretion cannot preferentially remove gas. Small dust is well coupled and must be carried along with the gas. Radial drift, in fact, may result in the rapid loss of mm-size grains where most of the solids mass is initially contained (e.g., Testi et al. 2014).   If larger planetesimals form prior to dispersal, it is not clear that they survive migration. Hasegawa \& Ida (2013) conclude that even gas giants may not survive migration in a massive disk. In a recent study, Coleman and Nelson (2014) modeled migration and planet formation via oligarchic growth (high relative planetesimal velocities) and dynamical evolution. They find that as long as the gas disk is present, the formation and retention of a giant planet ($\gtrsim 10$M$_{\oplus}$) is difficult because it would rapidly migrate. In their scenario, low mass disks form close-packed systems, and in high mass disks planets continuously form and migrate into the star and the last generation survives. All of these processes deplete solids relative to gas, and hence do not provide an explanation for the preferential removal of gas, at least in our solar system.
On the other hand, gas in photoevaporative flows is not dense enough to lift any dust but the smallest particles; since the mass is typically concentrated in larger sized particles this mechanism leaves most of the dust behind.

\subsection{Disk dispersal in the classic core accretion scenario}
Dispersal of protoplanetary disks by photo-evaporation has a strong effect on planet formation since it limits the time needed to build planetary systems. We recall that planet formation in the core accretion scenario (Lissauer \& Stevenson 2007) should explain a growth process through 12 orders of magnitude from the micron sized dust to Jupiter-like giant planets within the very limited lifetime of a protoplanetary disk. Planet formation in the core accretion scenario has different, well separated phases: (i) dust coagulation and formation of planetesimals, (ii) formation of planetary embryos and growth of the solid cores of giant planets, (iii) runaway gas accretion by the solid cores to form giant planets, and finally, (iv) the assembly of the planetary embryos to terrestrial planets. In the following we overview the timescales of the above phases of planet formation, and investigate how they are influenced by  photoevaporation of the disk. 
 
Formation of planetesimals due to dust coagulation is presently very uncertain due to several barriers to planetesimal growth: the bouncing barrier (Zsom et al. 2010), the charge barrier (Okuzumi et al. 2009), and the meter-size barrier (due to drift and fragmentation) (Blum \& Wurm 2000).  New approaches, such as formation in pressure maxima (Lyra et al. 2009), particle concentration due to streaming instability and gravoturbulent planetesimal formation (Johansen et al. 2014), have attempted to clarify these issues. In spite of many uncertainties, the timescale of particle growth to km-sized planetesimals is believed to be quite short; thus planetesimal formation takes place in the gas-rich protoplanetary disk and is unaffected by the disk's photoevaporation.

The next step is the further growth of planetesimals toward terrestrial planets leading to  runaway (Wetherill \& Stewart 1989) and oligarchic growth (Kokubo \& Ida 1998). These initial stages of the early phase of terrestrial planet formation in the classic core-accretion paradigm are rapid, lasting between 0.01 and 1 Myr. As the result of these processes, $\sim 100-1000$ bodies between Moon and Mars size are formed in the relevant region of terrestrial planet formation still in a gaseous environment. Planetesimals are affected by aerodynamic drag (due to the slightly sub-Keplerian motion of gas), while the formed oligarch feel torques leading to \emph{type I migration}. Drag and torques both result in angular momentum loss for the formed bodies, which spiral in.  Large planetesimals ($\gtrsim 100$ km in size) are less sensitive to drag force, but gravitationally perturb the ambient disk creating dense spiral wakes. Gravity due to these over-dense regions results in a net torque which modifies the planetesimal orbit. In earlier studies assuming isothermal disk models the net torque calculated was negative (Ward 1997) causing a significant loss of angular momentum and therefore inward migration of the planetesimal. More recent studies (Paardekooper et al. 2010) suggest the possibility of outward migration as well. We note that the speed of  type I migration is linearly proportional to the mass of the migrating body, therefore the rapid inward or outward migration of  relative massive objects will result in their loss from the region of terrestrial planet formation. 

Formation of giant planets clearly occurs in the presence of gas. Therefore, one of the most pressing issues in planet formation is building a giant planet within the few Myr disk lifetime. According to core accretion theory, first a solid core forms beyond the water snowline, where the abundance of solids is increased due to ice condensation on dust grains. The higher surface density of solids leads to efficient oligarchic growth allowing fast formation of a planetary core. The characteristic mass of a solid core able to capture a gaseous atmosphere at Jupiter's orbit is roughly $M_\mathrm{crit} = 1M_\oplus$. Both the core and envelope slowly grow, until the core reaches a critical mass of $\sim 10 M_{\oplus}$ and runaway gas accretion ensues leading to the rapid formation of a giant planet (on a $10^5$ year timescale). This rapid gas accretion slows down when the giant planet opens a gap in the gas disk. According to hydrodynamic simulations (Kley 1999) the planet is not entirely isolated from the ambient gas, and accretion continues along tidally generated spiral arms (Lubow \& D'Angelo 2006). Therefore, gas accretion onto the planet's surface stops only when the surrounding gas is dispersed and formation of a giant planet ends with the photoevaporation of the gas disk. 

In summary,  the presence of gas significantly influences the early phase of terrestrial planet formation; the effect of migration diminishes if much of the disk gas is dispersed before the runaway and oligarchic growth phases. Giant planets can form only if gas is present. However, the final assembly of terrestrial planets takes place in a few tens of million years, thus certainly in a gas free environment with no migration. We note that in more recent studies terrestrial planet formation may happen at \emph{planet traps} (places where the torque felt by a planet becomes zero) on much shorter timescales, well before the disk's final dispersal. In such models the effect of gas disk dispersal should be taken into account. These models will be briefly described in the next subsection.

\subsection{Disk dispersal and type I migration of terrestrial planets and planetary cores}
Planets and massive cores, after they form, can migrate, and type I migration can be very fast resulting in a rapid loss of the formed cores or planets. Planet traps can halt migration: traps can be caused by a change in disk thermal properties due to sudden changes in dust opacity (Lyra et al. 2010), or by the formation of a large scale vortex at the outer edge of the dead zone, which stops or reduces the migration speed of massive planetary cores (Regaly et al. 2013).

\begin{figure}[]
\includegraphics[scale=.4,angle=0]{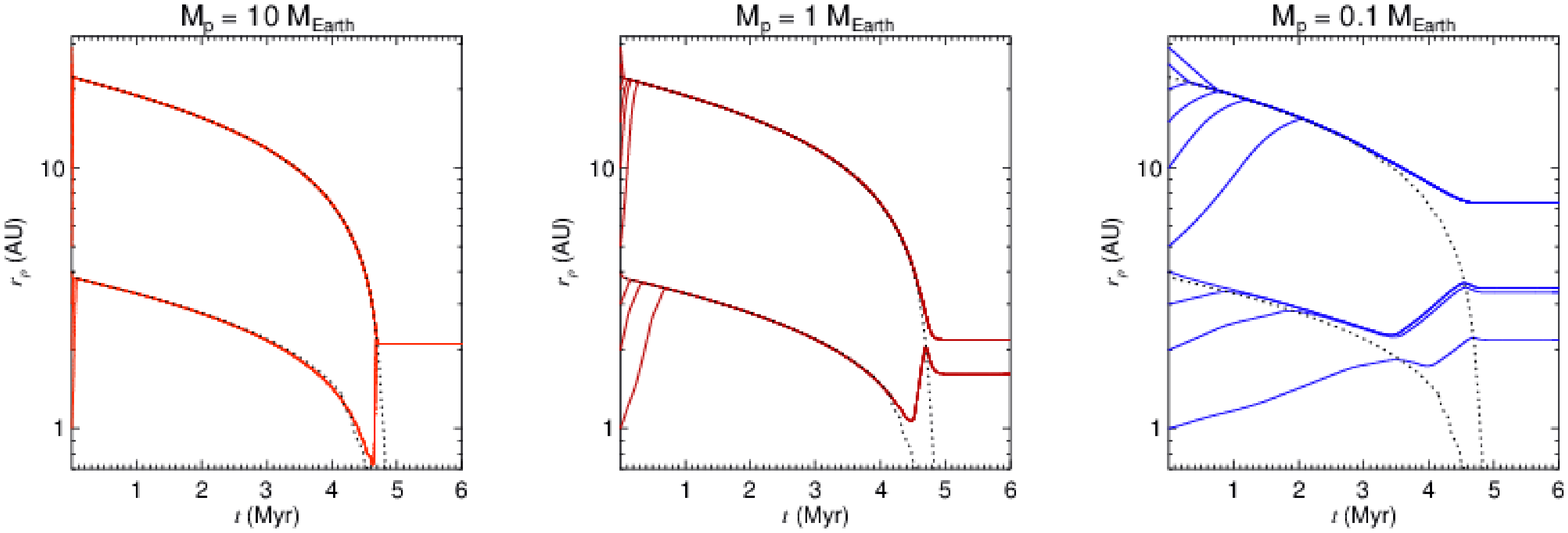}
\caption{Migration of different mass planets captured in the two planet traps in the model of Lyra et al. (2010) The orbital distances of planets are displayed with solid lines, while the dotted lines show the locations of the planet traps as functions of time. (Figure by courtesy of W. Lyra)}
\label{fig:planets_lyra}       
\end{figure}

Lyra et al. (2010) showed that opacity and temperature jumps in the disk can help prevent migration of planets (with $M_p\sim 0.1-10 M_\oplus$) using an evolving radiative 1D disk model with photoevaporation. In this model, the external photoevaporative disk wind prescription of Veras \& Armitage (2004) is used, the wind is effective only outside a critical radius of 5 AU.  The locations of the equilibrium radii with zero torque migrate due to evolution of the disk and the planet migrates toward these zero torque regions. As $\Sigma$ decreases, the radii of the zero torque locations (the traps) move faster toward the star than the migrating planets. As a consequence the planets are decoupled from equilibrium radii. $\Sigma$ at these times is too low to cause further migration. This effect is shown in Figure \ref{fig:planets_lyra}: in these simulations the evolution of the disk due to the combined effects of gas accretion and photoevaporation keep terrestrial planets from migrating into the star. 

\subsection{Effect of the disk dispersal on the type II migration of giant planets}
A massive Jupiter-like giant planet interacts with the disk to open an annular gap depleted of material, and then migrates on nearly viscous timescales (\emph{type II migration}, see chapter by Lin et al., this volume).  This migration timescale can be written as 
 (see e.g. in Baruteau \& Masset 2013):
\begin{equation}
\tau_\mathrm{II} = \frac{2 r_0^2}{3\nu(r_0)} \left( 1 + \frac{M_\mathrm{p}}{4\pi\Sigma(r_0) r^2} \right),
\label{eq:type2_formula}
\end{equation}
where $r_0 = r_\mathrm{p} + 2.5 r_\mathrm{Hill}$, and $r_\mathrm{Hill}$ is the Hill radius. The first term in the above formula is the effect of gas accretion and the second term is the ratio between the planet's mass and the local disk mass and can be interpreted as the ``resistance" of the inner disk to the inward migrating giant planet. In equilibrium, the gap moves at the accretion velocity during type II migration.  

Recent hydrodynamical simulations by Durmann \& Kley (2015) show that the migration of the giant planet is determined by the torques exerted by the disk. In general, they find that the migration of the giant planet does not follow the disk's viscous evolution and gas can flow through the gap. An important result of this research that if  $M_\mathrm{D}/M_\mathrm{P} < 0.2$ (where $M_\mathrm{D} = \Sigma(r_\mathrm{p})r_\mathrm{p}^2$ is interpreted as the local disk mass) the migration speed becomes significantly lower than the viscous speed (the type II migration rate). (We note that this behavior is also reflected in Equation \ref{eq:type2_formula} which accounts for the disk mass). While there is no simple analytical formulation for torque-based migration, these slower rates are  more consistent with planet synthesis models that can reproduce observations. Since a low disk mass is necessary to slow migration, the disk mass must be reduced at this stage by  the combined effect of gas accretion and photoevaporation. 

The final location of a giant planet is determined by the migration and disk dispersal times. Ignoring the effects of gravitational scattering in multi-planet systems, one can estimate the distribution of the semi-major axes of the randomly formed giant planets in a given disk model.
Alexander \& Armitage (2009) investigated the effects of gas accretion, EUV photo-evaporation, and the migration of the giant planet with a gap opening criterion (as in Lin \& Papaloizou 1986).  With plausible disk conditions and a range of planet masses ($0.5 M_\mathrm{Jup} < M_\mathrm{p} < 5 M_\mathrm{Jup}$), the orbital distribution of planets were found to be comparable to data from the Lick radial velocity suvey(Fischer \& Valenti 2005);  the two distributions are qualitatively similar (see Alexander \& Armitage 2009). Considering the effects of gap opening due to EUV photoevaporation, Alexander \& Pascucci (2012) found more recently that there would be deficit of planets at semi-major axis values close to the gap radius at $r_\mathrm{p} = 1-3$AU can be seen. This deficit is accompanied by a corresponding increase in the number of planets just outside these radii (see figure 2 of Alexander \& Pascucci 2012). We note, however, that  model results could be affected by many uncertainties due to migration rates which are linked to unknown aspects of planet formation (see discussion in Ercolano \& Rosotti 2015).

\subsection{Rapid photoevaporation of a disk with a giant planet} 
The formation of planets may in turn influence or accelerate disk dispersal. Conventional photoevaporation theory relies on the formation of a gap and then a hole, 
whose rim is irradiated directly to enhance mass loss subsequently. If a planet forms a gap and creates a similar rim, photoevaporation may be accelerated and trigger disk clearing (Alexander \& Armitage 2009). In a recent investigation (Rosotti et al. 2013) used the hydrodynamic code FARGO (Masset et al. 2000) coupled with a 1D code used for the initial $\sim 2$Myr evolution of the disk, including X-ray photoevaporation: these models yield a rapid dispersal of the inner disk (interior to the planet) compared to the case without X-ray photoevaporation where the inner disk persisted for the duration of the simulation. 

\section{Observational Constraints}
We next examine observational insights into the process of disk dissipation. Ideally, we would like to be able measure gas and dust masses, understand how their accretion rates decline, measure photoevaporative mass loss rates,  detect disks on the verge of dispersal, and assess their planet formation activity.    Rapid strides are being made in this area especially with newer, sensitive facilities capable of high spatial and spectral resolution. The broad scenario drawn from observations is consistent with the dispersal theory outlined so far, although details are not well understood. 
We first summarize what we currently know from observations and then discuss each of these: 
\begin{itemize}
\item Dust and gas disk lifetimes are $\lesssim 10 $Myr, very few disks survive beyond this timescale.
\item Disks evolve through a transition phase where the infrared excesses decrease; the inner disk probably clears first in most cases to form dust cavities. 
\item Photoevaporative winds have been detected in [NeII]$12.8\mu$m emission, and possibly also in [OI] forbidden line and CO emission; mass loss rates are yet to be determined. 
\item Exoplanet studies indicate that gas is present at late epochs of planet formation in most disks.
\end{itemize}

\subsection{Disk lifetimes} As discussed in \S 1, infrared excesses in disks appear to decline with an $e-$folding time of $\sim 3$Myr (e.g.,Mamajek 2009). Inner gas also disappears on similar timescales (Fedele et al. 2010), as is to be expected---gas if present would drag along small dust which causes the NIR excess. 
Moreover, Ribas et al. (2014) find that the fraction of sources with excesses increases at longer wavelengths, suggesting that the disk evolves inside-out. Inferred lifetimes are $\sim4-6$ Myr at $24\mu$m compared to $\sim2-3$ Myr at $3-12\mu$m.  The disk/star flux ratio at 24$\mu$m shows a sharp change at $\sim$10 Myr, suggesting that the nature of dust perhaps changes at these ages (Ribas et al. 2014).  There is a peak in the fraction of evolved disks (lower disk/star 24um flux) at $\sim10$ Myr, marking the transition from primordial to debris disk stage and the dispersal of gas (cf. Ercolano et al. 2011, Koepferl et al. 2015 for a theoretical perspective). Similarly, Wyatt (2008) argues that the disk mass derived from the sub-mm remains more or less constant (albeit with a wide dispersion, see Carpenter et al. 2014) and shows a sharp decline at 10 Myr, perhaps again indicating this transition to the debris disk phase (see also Wyatt et al. 2014). Since debris disks are almost always gas-free, the 10 Myr time also serves as an upper limit on the gas disk dispersal time, which is also consistent with gas observations (Zuckerman et al. 1995, Dent et al. 2013).

Photoevaporation can explain the above general behavior of dissipating disks quite well, and timescales are roughly in accordance with observations.  The inside-out dispersal of disks observed is, in fact, a main predicted characteristic of XEUV photoevaporation theory.  Although FUV photoevaporation causes mass loss in the outer disk, gap opening in the inner disk (at larger radii, $\sim 10$AU) typically precedes eventual disk dispersal here as well (Gorti et al. 2015).  The $2-10$ Myr timescales inferred above indicate photoevaporative mass loss rates that are at least of the order $\sim 10^{-9}$ to $ 10^{-8}$\mpy, in reasonable agreement with theoretical rates as discussed in \S 3 (Eq.~\ref{taudisk}). 

The sudden and simultaneous removal of dust along with gas is harder to explain. Alexander \& Armitage (2007) propose that as photoevaporation sweeps across the disk to remove gas, it may cause planetesimal formation at the expanding inner rim and deplete dust. This conversion needs to be rapid and highly efficient, if not, substantial amounts of dust may remain after gas disk removal. Debris disk processes such as PR drag could  remove the remnant primordial dust (see chapter by Wyatt et al. ), but these mechanisms act on Myr timescales and are not consistent with the rapid transition to the debris disk stage (e.g., Luhman et al. 2010, Wyatt et al. 2014). The most likely scenario is that planet formation has already removed most of the dust before gas disk dispersal, although this suggests a causal link between these two processes (see discussion in Gorti et al. 2015). 

\subsection{Transition Disks}
Transition disks are believed to represent one of possibly multiple pathways from primordial to debris disks (e.g. Williams \& Cieza 2011). Hence, this special class of objects are particularly relevant for disk dispersal theories.  While we conceptually understand how disks evolve and disperse, the origin and nature of transition disks is still under considerable debate.  (There are also several definitions of what constitutes a transition disk, we adopt the definition of Espaillat et al. (2014), i.e., disks with a clear deficit in short wavelength emission.) Transition disks have larger dust grains (Pinilla et al. 2014), are typically millimeter-bright (suggesting high disk mass) and accrete at rates a factor of $\sim 3-10$ lower than full primordial disks (e.g. Espaillat et al. 2014, Najita et al. 2015).  They also show differences in their gas emission, with
higher line ratios of HCN/H$_2$O in the infrared (Najita et al. 2013) and lower [O I] $63\mu$m line luminosities (Howard et el. 2012), trends that have both been explained as a result of their evolved dust content. 

The two main mechanisms proposed to explain transition disks are photoevaporation and planet-disk interactions.  XEUV photoevaporation predicts that about $\sim 10$\% of disks should be caught in the act of viscously draining their inner gas after gap formation (e.g., Owen et al. 2010), in agreement with the observed fraction of transition disks. However, photo evaporating disks are also predicted to be low in mass (note that this is the gas mass, however that it is the dust mass that is measured). FUV-dominated photoevaporation can open gaps at higher disk masses, but does not predict a large fraction of observable disks in the viscous draining phase ($\sim 10^4-10^5$ yr) because of longer dispersal times ($\sim 3-5$Myr). Gaps opened by planets in the cavities of transition disks is a more popular explanation, but not without difficulties. On the one hand, the planet needs to be massive to open a gap, $\gtrsim 1$M$_J$, and massive disks may be required to form giant planets, naturally explaining the higher mass of transition disks (Najita et al. 2015). On the other hand, Jupiters are believed to be rare (Winn \& Fabrycky 2014), and moreover, the formation time of giant planets is on the order of the disk dispersal time (e.g., D'Angelo et al. 2010), making the likelihood of observing an embedded Jovian mass planet too low to explain the $\sim 10$\% frequency of transition disks.  Multiple planetary systems are more common, but loss to migration and maintaining the observed stellar accretion rate past several planets pose problems (Zhu et al. 2011, Coleman \& Nelson 2015). Owen \& Clarke (2012) propose that there are two classes of transition disks, low mass disks compatible with photoevaporation theory and higher mass ones perhaps better explained by planet-disk interactions.  Rosotti et al. (2013) further consider the interaction between planet formation and photoevaporation and suggest that planets when they form could accelerate disk dispersal.  

While transition disks are believed to be an evolutionary stage that most disks go through, we note that a significant number of objects have, in fact, turned out to be unresolved binaries (e.g., CoKu Tau/4, CS Cha, HD142527 among others). Future, high resolution observations using faclities such as ALMA may shed light on the true nature of some of these disks (e.g., van der Marel et al. 2015). 

\subsection{Gas diagnostics of photoevaporation}

Emission from winds is the best method to directly measure the mass loss rates and assess the efficiency with which photoevaporation can deplete gas. The most promising such detection is the blue-shifted emission in [Ne II]$12.8\mu$m line from slow, thermal winds seen  from a few objects (Herczeg et al. 2007, Pascucci \& Sterzik 2009). Neon can be ionized by EUV ($\sim 21$ eV photons) and X-rays ($\sim 1$keV), and gas temperatures need to be $\gtrsim 500$K to excite the observed line.  Line luminosities and profiles are well reproduced by photoevaporation models (Alexander 2008, Ercolano \& Owen 2010), but mass loss rates using [NeII] are hard to determine without knowledge of the ionization level of the gas. For EUV-heated, fully ionized winds, data is consistent with mass loss rates of $\sim 10^{-10}$\mpy, while partly neutral X-ray heated gas implies higher rates $\sim 10^{-9}{\rm\ to\ } 10^{-8}$ \mpy (Hollenbach \& Gorti 2009, Pascucci et al. 2012). Other low velocity wind tracers such as [OI] forbidden line emission and CO rovibrational emission are more difficult to interpret with contributions from multiple components (Rigliaco et al. 2013). For a more in-depth discussion, see Alexander et al. (2014). 

Additional constraints on the ionization in the disk come from free-free emission, indicated by cm-excesses in a few disks  (e.g., Pascucci et al. 2012, Owen et al. 2013). Recent studies by Pascucci et al. (2014) and Galvan-Madrid et al. (2014) indicate that the EUV luminosities in disks are low based on observed free-free emission fluxes. These studies also find that the observed [NeII] emission is too high to arise from EUV-ionized gas at the inferred EUV luminosities. Therefore they conclude that the [NeII] emission must trace a wind ionized by hard $\sim 1$ keV X-rays, indirectly implying higher mass loss rates---due to either a X-ray driven photoevaporative wind, or a FUV-driven wind that is partly ionized by X-rays. 

Tracers of FUV photoevaporation are more difficult. Since the flows are launched subsonically and are considerably cooler, emission (for e.g., CO rotational lines) is dominated by the base of the flow which is at higher densities.  The molecules further get dissociated higher up in the wind. The blue-shifts and asymmetries in the line profiles expected here are small for detection with current facilities and hard to disentangle from other non-Keplerian sources like turbulence (Gorti et al., in preparation).  With the higher resolution and sensitivity of full ALMA and probes of the higher surface layers such as the weak [C I] 609$\mu$m line becoming accessible (Tsukagoshi et al. 2015), these flows may be detected in the near future.

\subsection{Gas at late stages and planet formation}

Exoplanet properties indicate that gas is present at late stages of planet formation in most disks. The most direct evidence stems from the detection of Super-Earths or mini-Neptunes---planets with masses $\gtrsim 2-3$M$_E$ and gaseous envelopes (Winn \& Fabrycky 2014). Close-in systems detected by Kepler transit surveys require gas both for in-situ and migration theories of their formation (e.g., Laughlin \& Lissauer 2015). The eccentricities of these systems are low, indicating that there were at least small quantities of gas present after the giant impact stage of forming terrestrial planets. On the other hand, the paucity of gas giants and the low gas masses of the Super-Earths indicate that there could not have been too much gas present. In that case, the planetary cores would have accreted more of the disk gas to form gas giants. 

Exoplanet masses and compositions therefore indicate that while gas was present at the epochs of planet formation, it dispersed shortly thereafter. This is particularly true for gas giant planet formation, with the final mass of the giant planet closely linked to the gas dispersal time (e.g., Lissauer et al. 2009, Movshovitz et al. 2010, Rogers et al. 2011).  

Contrary to the all the observational evidence presented so far, gas appears to persist in at least some debris disks (e.g., HD 21997, Kospal et al. 2013), well past planet formation epochs.  Although it is still unknown if the gas is primordial in origin, it is worth noting that all of the debris disks with gas detected are A stars, pointing to either longer disk dispersal times for intermediate-mass stars (however, see Ribas et al. 2014), or a detection bias. 

\section{Future Directions}
To summarize, the evolution of protoplanetary disks is initially dominated by viscous accretion, but at later critical planet-forming epochs internal and external photoevaporation by high energy photons (UV and X-rays) dictate the radial distribution of disk gas with time. Photoevaporation sets gas disk lifetimes and through its influence on gas disk evolution can impact all stages of planet formation, from planetesimal growth to the formation of giant planets. Although we qualitatively understand how disks evolve and disperse, photoevaporation rates are still not measured. Determining the decrease in gas mass with time and quantifying disk mass loss rates are essential toward developing a comprehensive theory of disk evolution that includes accretion, planet formation and disk dispersal.  We end with a list of possible future directions that may help resolve many outstanding issues: 
\begin{itemize}
\item
Some of the biggest uncertainties pertain to the stellar high energy spectrum. The flux of the accretion-generated X-ray and UV components, the relative strengths of the soft and hard X-ray fluxes and the strength of the Lyman $\alpha$ contribution to the FUV flux, and the evolution of all of these with time (along with relation to other variables such as accretion rates, disk and stellar masses) are some of the less well characterized inputs that need further investigation.  

\item Tracers of subsonic flow are almost non-existent, they may be needed to actually determine mass loss rates. Ideally, a measure of the gas mass for disks of different ages is desirable to quantify the rate at which disks dissipate. Gas emission line observations probe the density and temperature structure which are important for setting flow conditions; emission line modeling further indirectly measures disk irradiation.  Future observations from high sensitivity facilities like ALMA will inform disk heating and cooling physics and help calibrate disk models. 

\item More sophisticated disk models, which treat gas and dust separately and include hydrodynamics, radiative transfer and chemistry are needed. With rapid advances in supercomputing facilities and techniques, such models may soon become possible. Disk evolution models need to self-consistently account for planet formation and dynamical evolution along with disk dispersal. As comprehensive studies become more common (e.g., Coleman \& Nelson 2015), future work may allow for advanced population synthesis models that can simultaneously explain the diversity in disk and exoplanet properties. 
 \end{itemize}

Ultimately, we would like to be able to connect disk evolution to planet formation and understand the close, and perhaps causal, correspondence between timescales for planet formation and disk dispersal.


\begin{acknowledgements}
We thank ISSI for their kind invitation and our Chinese hosts for their hospitality during the workshop. U.~Gorti acknowledges support from the National Science Foundation (NSF AST-1313003). Zs. S\'andor thanks the support of the J\'anos Bolyai Research Scholarship of the Hungarian Academy of Sciences. 
\end{acknowledgements}

\end{document}